\documentclass[aps,prl,showpacs]{revtex4}
\usepackage{amsmath}
\usepackage{graphicx}
\usepackage{bm}

\begin{document}

\title{Quaternionic Root Systems and Subgroups of the $Aut(F_{4})$}
\date{\today}
\author{Mehmet Koca}
\email{kocam@squ.edu.om}
\affiliation{Department of Physics, College of Science, Sultan Qaboos University, PO Box
36, Al-Khod 123, Muscat, Sultanate of Oman}
\author{Ramazan Ko\c{c}}
\email{koc@gantep.edu.tr}
\affiliation{Department of Physics, Faculty of Engineering University of Gaziantep, 27310
Gaziantep, Turkey}
\author{Muataz Al-Barwani}
\email{muataz@squ.edu.om}
\affiliation{Department of Physics, College of Science, Sultan Qaboos University, PO Box
36, Al-Khod 123, Muscat, Sultanate of Oman}

\begin{abstract}
Cayley-Dickson doubling procedure is used to construct the root systems of
some celebrated Lie algebras in terms of the integer elements of the
division algebras of real numbers, complex numbers, quaternions and
octonions. Starting with the roots and weights of $SU(2)$ expressed as the
real numbers one can construct the root systems of the Lie algebras of $%
SO(4),SP(2)\approx SO(5),SO(8),SO(9),F_{4}$ and $E_{8}$ in terms of the
discrete elements of the division algebras. The roots themselves display the
group structures besides the octonionic roots of $E_{8}$ which form a closed
octonion algebra. The automorphism group $Aut(F_{4})$ of the Dynkin diagram
of $F_{4}$ of order 2304, the largest crystallographic group in
4-dimensional Euclidean space, is realized as the direct product of two
binary octahedral group of quaternions preserving the quaternionic root
system of $F_{4}$ .The Weyl groups of many Lie algebras, such as, $%
G_{2},SO(7),SO(8),SO(9),SU(3)XSU(3)$ and $SP(3)\times SU(2)$ have been
constructed as the subgroups of $Aut(F_{4})$ .We have also classified the
other non-parabolic subgroups of $Aut(F_{4})$ which are not Weyl groups. Two
subgroups of orders192 with different conjugacy classes occur as maximal
subgroups in the finite subgroups of the Lie group $G_{2}$ of orders 12096
and 1344 and proves to be useful in their constructions. The triality of $%
SO(8)$ manifesting itself as the cyclic symmetry of the quaternionic
imaginary units $e_{1},e_{2},e_{3}$ is used to show that $SO(7)$ and $SO(9)$
can be embedded triply symmetric way in $SO(8)$ and $F_{4}$ respectively.
\end{abstract}

\maketitle

\section{Introduction}

There are a few celebrated Lie algebras which seem to be playing important
roles in understanding the underlying symmetries of the unified theory of
all interactions. The most popular ones are the exceptional Lie groups $%
G_{2},F_{4},E_{6},E_{7}$ and $E_{8}$ and the related groups \cite{1}. The
groups $Spin7$ and $G_{2}$ are proposed as the holonomy groups for the
compactification of the M-theory from 11 to 4 dimensional space-time \cite{2}%
. It is also well known that two orthogonal groups $SO(8)$ and $SO(9)$ are
the little groups of the massles particles of string theories in
10-dimensions and the M-theory in 11-dimensions respectively. The fact that $%
SO(9)$ can be embedded in the exceptional Lie group $F_{4}$ in a triply
symmetric way and the non-compact $F_{4(-25)}$ can be embedded in the Lorenz
group $SO(25,1)$ indicates the importance of the exceptional group $F_{4}$ %
\cite{3}. The largest exceptional group $E_{8}$ which had been suggested as
the unified theory of the electroweak and strong interactions with three
generations of lepton-quark families \cite{4} naturally occurred as the
gauge symmetry of the $E_{8}\times E_{8}$ heterotic string theory \cite{5}.
It has many novel mathematical aspects \cite{6} which has not been exploited
in physics. It was known that a non-compact version of $E_{7}$ manifests
itself as a global symmetry of the 11-dimensional supergravity \cite{7}.
Some of its maximal subgroups show themselves as local symmetries \cite{8}.
The $E_{6}$ has been suggested as a unified theory of electroweak and strong
interactions \cite{9}.

The Weyl groups of these groups are also important for the invariants of the
Lie groups can be deduced using the related Weyl groups. The Weyl groups of
the exceptional Lie groups $F_{4},E_{6},E_{7}$ and $E_{8}$ correspond to
some finite subgroups of the Lie groups of $O(4),O(6),O(7)$ and $O(8)$
respectively \cite{10}.

It has been shown in some details that the finite subgroups of $O(4)$ can be
classified as direct products of finite subgroups of quaternions \cite{11},
isomorphic to the finite subgroups of $SU(2)$ ,which is the double cover of $%
SO(3)$ .Therefore the relevant Weyl groups of the Lie groups $F_{4},SO(9)$
and $SO(8)$ correspond to some finite subgroups of $O(4)$ \cite{12}.
Similarly, the Weyl groups of some rank-3 Lie algebras can be obtained from
the finite subgroups of $O(3)$ . Interestingly enough, the relevant root
systems can be represented as discrete quaternions and the Weyl groups can
be realized as the left-right actions of the quaternions on the root
systems. When one considers the finite subgroups of $O(8)$ it is natural to
encounter with the discrete octonions which represent the root system of $%
E_{8}$ where the root system of $E_{7}$ is described by imaginary octonions %
\cite{13}. The automorphism group of octonionionic root system of $E_{7}$
turns out to be a finite subgroup of $G_{2}$ of order 12096 \cite{14}. In
what follows we will restrict ourselves to the quaternionic root system of $%
F_{4},SO(9),SO(8),SO(7),SP(3)$ and construct explicitly their Weyl groups as
finite subgroups of $O(4)$ .The largest group of interest here is the $%
Aut(F_{4})$ of order $2304=48\times 48$ which is the direct product of the
binary octahedral group with itself \cite{15}. We follow a chain of
decomposition of $Aut(F_{4})$ into its relevant subgroups, some of which,
are maximal subgroups in the finite subgroups of $G_{2}$ of orders 12096 and
1344 \cite{16}.

The paper is organized as follows. In section 2 we start with the scaled
roots $\pm 1,0$ and the weights $\pm \frac{1}{2}$ of $SU(2)$ and using the
Cayley-Dickson doubling procedure we construct the roots of $SO(4)$ and $%
SP(2)\approx SO(5)$ in terms of complex numbers. Further doubling of the
roots of $SP(2)\approx SO(5)$ leads to the quaternionic roots of $SO(8)$ %
\cite{13,16}. The 8-dimensional vector and spinor representations of Spin-8
constitute the short roots of $F_{4}$ . Doubling of two sets of quaternionic
roots of $F_{4}$ leads to the octonionic roots of $E_{8}$ . The triality of $%
SO(8)$ is then coded in the cyclic symmetry of the quaternionic imaginary
units. In section 3 we introduce the finite subgroups of $SU(2)$ in terms of
quaternions and explain their geometric properties. We explain how to
construct the $Aut(F_{4})$ and the Weyl groups of $F_{4},SO(9)$ and $SO(8)$ %
\cite{15}. In section 4 we construct the root systems of $SO(7)$ and $G_{2}$
by folding the Coxeter-Dynkin diagram of $SO(8)$ \cite{17} which displays
the three-fold embeddings of $SO(7)$ into $SO(8)$ .The Weyl groups of $SO(7)$
, $G_{2}$ and $SP(3)$ are constructed in terms of quaternions. In section 5
we discuss the subgroup chains of $Aut(F_{4})$ and find out the explicit
expressions of the groups down to the groups of order 192. A particular
emphasis is given to two groups of orders 192 since they appear as the
maximal subgroups in the finite subgroups of $G_{2}$ of orders 12096 and
1344. Finally in section 6 we further elaborate the geometric aspects of the
symmetries discussed in the previous sections.

\section{Root Systems with the Cayley-Dickson Doubling Procedure}

The Cayley- Dickson doubling is a procedure to build the elements of
division algebras starting with the real numbers. Let us denote by $p,q,r,s$
the elements of a division algebra other than the octonions. Then the pairs $%
(p,q)$ and $(r,s)$ with the multiplication rule%
\begin{equation}
(p,q)(r,s)=(pr-s\bar{q},rq+\bar{p}s)  \label{e1}
\end{equation}%
constitute the elements of a division algebra in higher dimension. The
celebrated Hurwitz's theorem[18] states that there are only four division
algebras, namely, real numbers, complex numbers, quaternions and octonions.
Starting with the complex numbers at every higher level of division algebras
one introduces one complex number, say, $e_{1},e_{2}$ and $e_{7}$ which
anti-commute with each other and satisfy the relation $%
e_{1}^{2}=e_{2}^{2}=e_{7}^{2}=-1$. Doubling of the real numbers constitutes
the complex numbers, two sets of complex numbers define the quaternions and
finally a pair of quaternions defines the octonions under the definition (%
\ref{e1}). Let us revise the work of reference \cite{13} by starting with
the roots $\pm 1,0$ and the weights $\pm \frac{1}{2}$ of $SU(2)$. A pair of
set $\pm 1,0$ leads to the roots of $SO(4)$:%
\begin{equation}
(\pm 1,0)=\pm 1,(0,\pm 1)=\pm e_{1}(\mathrm{we\quad use\quad }e_{1}\quad 
\mathrm{for\quad the\quad imaginary\quad number\quad }i)  \label{e2}
\end{equation}%
The non-zero roots $\pm 1,\pm e_{1}$ of $SO(4)$ form a cyclic group of order
4. The weights of the spinor representation $(\underline{2},\underline{2})$
of $SO(4)$ can be taken as%
\begin{equation}
\left( \pm \frac{1}{2},\pm \frac{1}{2}\right) =\frac{1}{2}\left( \pm 1\pm
e_{1}\right) .  \label{e3}
\end{equation}%
The roots in (\ref{e2}) and the weights in (\ref{e3}) constitute the scaled
roots of $SP(2)\approx SO(5)$,%
\begin{equation}
SP(2)\approx SO(5):\pm 1,\pm e_{1},\frac{1}{2}(\pm 1\pm e_{1}).  \label{e4}
\end{equation}

When the short roots are scaled to the unit norm then the roots of $SP(2)$
form a cyclic group of order 8. A non-trivial structure will arise when two
sets of the roots of $SP(2)$ are paired as $(SP(2),SP(2))$ where the long
roots match with the zero roots while the short roots match with the short
roots leading to the quaternionic roots of $SO(8)$ :%
\begin{equation}
T:\left\{ \pm 1,\pm e_{1},\pm e_{2},\pm e_{3},\frac{1}{2}(\pm 1\pm e_{1}\pm
e_{2}\pm e_{3})\right\}  \label{e5}
\end{equation}%
where we have used $e_{3}e_{1}=-e_{1}e_{3}=e_{2}$. If we include the pairing
of the short roots with the zero roots we obtain%
\begin{equation}
V_{1}^{\prime }:\left( \frac{1}{2}(\pm 1\pm e_{1}),0\right) =\frac{1}{2}(\pm
1\pm e_{1}),\left( 0,\frac{1}{2}(\pm 1\pm e_{1})\right) =\frac{1}{2}(\pm
e_{2}\pm e_{3})  \label{e6}
\end{equation}

These are the weights of the 8-dimensional representation of $SO(8)$ and
together with the roots in (\ref{e5}) they represent the roots of $SO(9)$.
The cyclic symmetry of the quaternionic imaginary units would lead to the
weights of the two 8-dimensional spinor representations of $SO(8)$ which
represent the weights of the 16-dimensional spinor representation of $SO(9)$
:%
\begin{equation}
\begin{array}{ccccc}
& \frac{1}{2}(\pm 1\pm e_{2}) &  &  & \frac{1}{2}(\pm 1\pm e_{3}) \\ 
V_{2}^{\prime }: & \frac{1}{2}(\pm e_{3}\pm e_{1}) &  & V_{3}^{\prime }: & 
\frac{1}{2}(\pm e_{1}\pm e_{2})%
\end{array}
\label{e7}
\end{equation}%
The set of quaternions in (\ref{e5}-\ref{e7}) constitutes the scaled roots
of $F_{4}$ . A further doubling the set of roots of $F_{4}$ will lead to the
octonionic roots of $E_{8}$ \cite{13}:%
\begin{eqnarray}
&(T,0)=&T,\quad (0,T)=e_{7}T  \notag \\
&(V_{1}^{\prime },V_{1}^{\prime })=&V_{1}^{\prime }+e_{7}V_{1}^{\prime } 
\notag \\
&(V_{2}^{\prime },V_{3}^{\prime })=&V_{2}^{\prime }+e_{7}V_{3}^{\prime }
\label{e8} \\
&(V_{3}^{\prime },V_{2}^{\prime })=&V_{3}^{\prime }+e_{7}V_{2}^{\prime } 
\notag
\end{eqnarray}%
where one can define $e_{4}=e_{7}e_{1}$ , $e_{5}=e_{7}e_{2}$ , $%
e_{6}=e_{7}e_{3}$ .We note that when the roots in (\ref{e6}-\ref{e7}) are
multiplied by $\sqrt{2}$ to make the norm $1$ then the 48 set of quaternions
are the elements of the binary octahedral group $O$ of $SU(2)$ where $T$
represents the binary tetrahedral subgroup of order 24.

\section{Binary Octahedral group and the $Aut(F_{4})$}

Some of the material of this section have been discussed in reference \cite%
{12}. The finite subgroups of $SO(3)$ are well known: icosahedral group of
order 60, octahedral group of order 24, tetrahedral group of order 12, and
dihedral and cyclic groups of various orders \cite{19}. Their double covers
are the finite subgroups of quaternions which are related to the ADE series
of the Lie algebras through the McKay correspondence \cite{20}. Our interest
here solely are constrained to the binary octahedral group whose direct
product with itself is isomorphic to the $Aut(F_{4})$ which can be realized
as the left and right actions of the quaternionic elements on the
quaternionic roots of $F_{4}$. The root system of $F_{4}$ has very
interesting geometrical structures which has not been discussed in the
literature. We classify the elements of the binary octahedral group as sets
of the hyperoctahedra in 4-dimensions \cite{21}:%
\begin{equation}
\begin{array}{cll}
& V_{0}=\left\{ {\pm 1,\pm e_{1},\pm e_{2},\pm e_{3}}\right\} &  \\ 
T: & V_{+}=\left\{ \frac{1}{2}{\pm 1\pm e_{1}\pm e_{2}\pm e_{3}}\right\} , & 
\mathrm{even\quad number\quad of\quad (+)\quad signs} \\ 
& V_{-}=\overline{V_{+}}=\left\{ \frac{1}{2}{\pm 1\pm e_{1}\pm e_{2}\pm e_{3}%
}\right\} {,} & \mathrm{even\quad number\quad of\quad (+)\quad signs}%
\end{array}
\label{e9a}
\end{equation}

where $\overline{V_{+}}$ is the quaternionic conjugate of $V_{+}.$ 
\begin{eqnarray}
&&%
\begin{array}{cl}
& V_{1}=\left\{ {\frac{1}{\sqrt{2}}(\pm 1\pm e_{1}),\frac{1}{\sqrt{2}}(\pm
e_{2}\pm e_{3})}\right\} \\ 
T^{\prime }: & \left\{ {\frac{1}{\sqrt{2}}(\pm 1\pm e_{2}),\frac{1}{\sqrt{2}}%
(\pm e_{3}\pm e_{1})}\right\} \\ 
& \left\{ {\frac{1}{\sqrt{2}}(\pm 1\pm e_{3}),\frac{1}{\sqrt{2}}(\pm
e_{1}\pm e_{2})}\right\}%
\end{array}
\label{e9b} \\
&&%
\begin{array}{cl}
O: & T\oplus T^{\prime }%
\end{array}
\label{e9c}
\end{eqnarray}

Here each of $V_{0}$ , $V_{+}$ and $V_{-}$ represents the vertices of a
hyperoctahedron in 4-dimensions and any two hyperoctahedra form a hypercube
in 4-dimensions with 16 vertices. The set of quaternions $T$ in (9a) not
only constitute the non-zero roots of $SO(8)$ but also represent a polytope $%
{3,4,3}$ called 24-cell \cite{21}. The set of quaternions in $T^{\prime }$
are the duals of $T$ ; consequently any $V_{i}(i=1,2,3)$ is a
hyperoctahedron and any two hyperoctahedra form the vertices of a hypercube.
We give the multiplication table of these sets of quaternions in Table1 to
understand the structure of the binary octahedral group. Here $V_{0}$ is the
quaternion group and form an invariant subgroup both in $T$ and $O$. 
\begin{table}[t]
$%
\begin{tabular}{|lllllll|}
\hline
& $V_{0}$ & $V_{+}$ & $V_{-}$ & $V_{1}$ & $V_{2}$ & $V_{3}$ \\ \hline
\multicolumn{1}{|l|}{$V_{0}$} & \multicolumn{1}{|l}{$V_{0}$} & $V_{+}$ & 
\multicolumn{1}{l|}{$V_{-}$} & $V_{1}$ & $V_{2}$ & $V_{3}$ \\ 
\multicolumn{1}{|l|}{$V_{+}$} & \multicolumn{1}{|l}{$V_{+}$} & $V_{-}$ & 
\multicolumn{1}{l|}{$V_{0}$} & $V_{3}$ & $V_{1}$ & $V_{2}$ \\ 
\multicolumn{1}{|l|}{$V_{-}$} & \multicolumn{1}{|l}{$V_{-}$} & $V_{0}$ & 
\multicolumn{1}{l|}{$V_{+}$} & $V_{2}$ & $V_{3}$ & $V_{1}$ \\ \cline{2-7}
\multicolumn{1}{|l|}{$V_{1}$} & \multicolumn{1}{|l}{$V_{1}$} & $V_{2}$ & 
\multicolumn{1}{l|}{$V_{3}$} & $V_{0}$ & $V_{+}$ & $V_{-}$ \\ 
\multicolumn{1}{|l|}{$V_{2}$} & \multicolumn{1}{|l}{$V_{2}$} & $V_{3}$ & 
\multicolumn{1}{l|}{$V_{1}$} & $V_{-}$ & $V_{0}$ & $V_{+}$ \\ 
\multicolumn{1}{|l|}{$V_{3}$} & \multicolumn{1}{|l}{$V_{3}$} & $V_{1}$ & 
\multicolumn{1}{l|}{$V_{2}$} & $V_{+}$ & $V_{-}$ & $V_{0}$ \\ \hline
\end{tabular}%
$%
\caption{Multiplication table of the binary octahedral group}
\label{tab:b}
\end{table}

A general element of $O(4)\approx SU(2)\times SU(2)$ can be defined as
follows. Denote by $p,q$ the quaternions of unit norm acting on an arbitrary
quaternion $r=r_{0}+r_{1}e_{1}+r_{2}e_{2}+r_{3}e_{3}$%
\begin{eqnarray}
^{\text{\cite{not1}}}r &\rightarrow &prq:[p,q]  \label{e10a} \\
r &\rightarrow &p\bar{r}q:[p,q]^{\ast }  \label{e10b}
\end{eqnarray}%
where $\bar{r}$ is the quaternion conjugate $\bar{r}%
=r_{0}-r_{1}e_{1}-r_{2}e_{2}-r_{3}e_{3}$ . For arbitrary quaternions $p,q$
with unit norm the elements $[p,q]$ and $[p,q]^{\ast }$ form a six parameter
group leaving the norm $r\bar{r}=\bar{r}r$ invariant. When written in terms
of matrices the group elements $[p,q]$ and $[p,q]^{\ast }$ have determinants 
$+1$ and $-1$ respectively. Therefore the elements $[p,q]$ form a subgroup $%
SO(4)\approx \frac{SU(2)xSU(2)}{Z_{2}}$ of $O(4)$ . In the reference \cite%
{12} we have proven that the Weyl group $W(F_{4})$ can be compactly written
as the union of elements,%
\begin{equation}
W(F_{4})={[T,T]\oplus \lbrack T\prime ,T\prime ]\oplus \lbrack T,T]^{\ast
}\oplus \lbrack T\prime ,T\prime ]^{\ast }}  \label{e11}
\end{equation}%
The automorphism group $Aut(F_{4})$ is the semi-direct product of the Weyl
group $W(F_{4})$ with the $Z_{2}$ symmetry of the Coxeter-Dynkin diagram of $%
F_{4}$,%
\begin{equation}
^{\text{\cite{not2}}}Aut(F_{4})\equiv {(O,O)\oplus (O,O)^{\ast }}\approx
W(F_{4}):Z_{2}.  \label{e12}
\end{equation}%
The generators of $Aut(F_{4})$ can be obtained from the Coxeter-Dynkin
diagram of $F_{4}$ where the simple roots are given in terms of scaled
quaternions.

\begin{figure}[h]
\begin{center}
\begin{picture}(300,120)(0,0)
\put(45,15){$\alpha _{4}^{\prime }=\frac{1}{2}(e_{1}-e_{2})$}
\put(120,15){\circle*{5}}
\put(120,15){\line(0,60){30}}
\put(125,45){$\alpha _{3}^{\prime }=\frac{1}{2}(e_{2}-e_{3})$}
\put(120,45){\circle*{5}}
\put(118,45){\line(0,60){30}}
\put(122,45){\line(0,60){30}}
\put(75,75){$\alpha _{2}^{\prime }=e_{3}$}
\put(120,75){\circle{5}}
\put(120,73){\line(0,60){30}}
\put(125,105){$\alpha _{1}^{\prime }=\frac{1}{2}(1-e_{1}-e_{2}-e_{3})$}
\put(120,105){\circle{5}}
\end{picture}
\end{center}
\caption{The Coxeter-Dynkin diagram of $F_{4}.$}
\label{fig:one}
\end{figure}
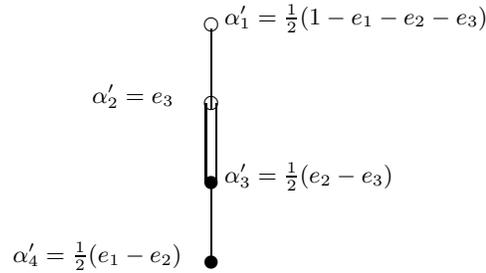

The regular simple roots $\alpha _{i}$ are related to $\alpha _{i}^{\prime }$
by $\alpha _{i}=\sqrt{2}\alpha _{i}^{\prime }(i=1,2,3,4)$. The $Aut(F_{4})$
is generated by the elements%
\begin{equation}
\lbrack \alpha _{1}^{\prime },-\alpha _{1}^{\prime }]^{\ast },[\alpha
_{2}^{\prime },-\alpha _{2}^{\prime }]^{\ast },[\alpha _{3}^{\prime
},-\alpha _{3}^{\prime }]^{\ast },[\alpha _{4}^{\prime },-\alpha
_{4}^{\prime }]^{\ast },[\frac{1}{\sqrt{2}}(e_{2}+e_{3}),-e_{2}]  \label{e13}
\end{equation}

The first four generators in (\ref{e13}) represent the reflections in the
roots $\alpha _{i}(i=1,2,3,4)$ and generate the Weyl group $W(F_{4})$ and
the last term stands for the diagram symmetry of $F_{4}$ which transforms,
by conjugation, $\alpha _{1}^{\prime }\leftrightarrow \alpha _{4}^{\prime }$
and $\alpha _{2}^{\prime }\leftrightarrow \alpha _{3}^{\prime }$. An
extended Coxeter-Dynkin diagram of $F_{4}$ can be used to obtain the
Coxeter-Dynkin diagrams of its maximal Lie algebras. We will discuss all
starting with $SO(9)$ .

\subsection{The Parabolic Subgroups of $F_{4}$}

\subsubsection{$SO(9)$}

We have shown in reference \cite{12} that the Weyl group $W(SO(9))$ can be
represented by the set of group elements%
\begin{eqnarray}
&&\lbrack V_{0},V_{0}],[V_{+},V_{+}],[V_{-},V_{-}],[V_{0},V_{0}]^{\ast
},[V_{+},V_{+}]^{\ast },[V_{-},V_{-}]^{\ast }  \label{e14a} \\
&&\lbrack V_{1},V_{1}],[V_{2},V_{3}],[V_{3},V_{2}],[V_{1},V_{1}]^{\ast
},[V_{2},V_{3}]^{\ast },[V_{3},V_{2}]^{\ast }  \label{e14b}
\end{eqnarray}

This is a group of order 384. The $W(SO(9))$ can be embedded in the $%
W(F_{4}) $ triply symmetric way by permuting the quaternionic imaginary
units $e_{1},e_{2},e_{3}$ in the cyclic order. It is an inner automorphism
of $W(F_{4})$ which replaces the elements in (\ref{e14b}) by the
corresponding elements where the indices1,2,3 are permuted in the cyclic
order. This permutation of the indices leaves the set of elements in (\ref%
{e14a}) invariant as expected. Actually the set of elements in (\ref{e14a})
constitute the elements of the Weyl group $W(SO(8))$. The Weyl group $%
W(SO(9))$ has a very interesting geometrical aspect; it is the largest
symmetry preserving the 4-dimensional hyperoctahedron. One can show that the
group elements in (\ref{e14a}-\ref{e14b}) leave the set of elements in $%
V_{1} $ invariant which is one of those 6 hyperoctahedra of the 48 roots of $%
F_{4}$. This has to be expected anyway because the set of roots $V_{1}/\sqrt{%
2}$ are the short roots of $SO(9)$ and has to be rotated to each other by
the elements of $W(SO(9))$. Since the weights of the 16-dimensional spinor
representation are represented by the quaternions $\frac{1}{\sqrt{2}}%
(V_{2}+V_{3})$ corresponding to the vertices of a cube in 4-dimensions they
are also preserved by the elements of $W(SO(9))$ in (14a-b). Embedding $%
W(SO(9))$ in $Aut(F_{4})$ can be made with a six fold cyclic symmetry under
the conjugation, say, by $[V_{+},V_{1}]W(SO(9))[V_{-},V_{1}]$ where $%
[V_{+},V_{1}]^{6}=[V_{0},V_{0}]$ . This leads to six conjugate
representations of $W(SO(9))$ in $Aut(F_{4})$ in each of which one of the
six hyperoctahedra $V_{0},V_{\pm },V_{i}(i=1,2,3)$ is left invariant by $%
W(SO(9))$ .There are other subgroups of $Aut(F_{4})$ of order 384 not
isomorphic to the Weyl group $W(SO(9))$. We will discuss them in section 5.
Now we discuss the Weyl group of the maximal subalgebra $SU(2)\times SP(3)$
of $F_{4}$.

\subsubsection{$SU(2)\times SP(3)$}

The algebra $SU(2)\times SP(3)$ can be represented by the Coxeter-Dynkin
diagram shown in figure 2.

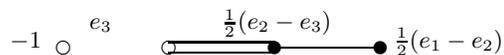
\begin{figure}[h]
\begin{center}
\begin{picture}(300,120)(0,0)
\put(100,15){$-1$}
\put(120,15){\circle{5}}
\put(130,22){$e_{3} $}
\put(160,15){\circle{5}}
\put(160,13){\line(60,0){40}}
\put(160,17){\line(60,0){40}}
\put(180,22){$\frac{1}{2} (e_{2} -e_{3} )$}
\put(200,15){\circle*{5}}
\put(200,15){\line(60,0){40}}
\put(245,15){$\frac{1}{2} (e_{1} -e_{2} )$}
\put(240,15){\circle*{5}}
\end{picture}
\end{center}
\caption{The Coxeter-Dynkin diagram of $SU(2)\times SP(3)$.}
\label{fig:two}
\end{figure}

The reflection generators on the simple roots are represented by%
\begin{equation}
r_{0}=[1,-1]^{\ast },r_{1}=[e_{3},-e_{3}]^{\ast },r_{2}=[\frac{1}{\sqrt{2}}%
(e_{2}-e_{3}),-\frac{1}{\sqrt{2}}(e_{2}-e_{3})]^{\ast },r_{3}=[\frac{1}{%
\sqrt{2}}(e_{1}-e_{2}),-\frac{1}{\sqrt{2}}(e_{1}-e_{2})]^{\ast }  \label{e15}
\end{equation}

and will generate the set of roots%
\begin{equation}
\begin{array}{cc}
SU(2) & SP(3) \\ 
\pm 1 & \pm e_{1},\pm e_{2},\pm e_{3} \\ 
& \frac{1}{2}(\pm e_{1}\pm e_{2}) \\ 
& \frac{1}{2}(\pm e_{2}\pm e_{3}) \\ 
& \frac{1}{2}(\pm e_{3}\pm e_{1})%
\end{array}
\label{e16}
\end{equation}

The long roots $\pm e_{1},\pm e_{2},\pm e_{3}$ of $SP(3)$ form the vertices
of an octahedron. Therefore the Weyl group $W(SP(3))$ is the symmetry of the
octahedron in 3-dimension. Since the product of two reflections is a
rotation around some axis the proper rotation subgroup of $W(SP(3))$ is
generated by%
\begin{equation}
R=r_{1}r_{2}=[\frac{1}{\sqrt{2}}(1-e_{1}),\frac{1}{\sqrt{2}}%
(1+e_{1})],S=r_{2}r_{3}=[\bar{t},t]  \label{e17}
\end{equation}%
with $t=\frac{1}{2}(1+e_{1}+e_{2}+e_{3})$. Here the generators satisfy the
generation relations of an octahedral group \cite{23}%
\begin{equation}
R^{4}=S^{3}=(RS)^{2}=[1,1].  \label{e18}
\end{equation}

It is one of the finite subgroups of $SO(3)$ isomorphic to the symmetric
group $S_{4}$. Another generator $(r_{1}r_{2}r_{3})^{3}=[1,1]^{\ast }$
commutes with the generators $R$ and $S$ so that the maximal group of the $%
SP(3)$ roots is the group $W(SP(3))\approx S_{4}\times Z_{2}$, a group of
order 48. The $Z_{2}$ group of $W(SU(2))$ is generated by $[1,-1]^{\ast }$
which commutes with the generators of $W(SP(3))$. Therefore the Weyl group $%
W(SU(2))xW(SP(3))$ is isomorphic to the group $S_{4}\times Z_{2}^{2}$ of
order 96. The group elements are represented by the pair of quaternions%
\begin{equation}
\lbrack p,\pm \bar{p}],[p^{\prime },\pm \bar{p}],[p,\pm \bar{p}]^{\ast
},[p^{\prime },\pm \bar{p}];p\in T,p^{\prime }\in T^{\prime }.  \label{e19}
\end{equation}

Since the vertices of the octahedron are represented by the imaginary
quaternions $\pm e_{1},\pm e_{2},\pm e_{3}$ one can naturally ask the
question: what is the maximal group which preserves the quaternion algebra
of the set of quaternions $\pm e_{1},\pm e_{2},\pm e_{3}$? It is well known
that when $p$ is the unit quaternion with non-zero real component then the
transformation $e\prime _{i}=pe_{i}\bar{p}$ is the only transformation which
preserves the quaternion algebra and is isomorphic to the group $SO(3)$.
This implies that the finite subgroup of $SO(3)$ which preserves the set of
quaternions $\pm e_{1},\pm e_{2},\pm e_{3}$ is the octahedral group
represented by the elements $[p,\bar{p}],[p\prime ,\bar{p}]$ which is
isomorphic to the symmetric group $S_{4}$.

\subsubsection{$SU(3)\times SU(3)$}

From the extended Coxeter-Dynkin diagram of $F_{4}$ we obtain that the
Coxeter-Dynkin diagram of $SU(3)xSU(3)$:

\begin{figure}[h]
\begin{center}
\begin{picture}(300,120)(0,0)
\put(58,19){$-1$}
\put(60,15){\circle{5}}
\put(61,15){\line(60,0){38}}
\put(98,19){$\bar{t} $}
\put(100,15){\circle{5}}
\put(138,19){$s_{1}^{\prime }$}
\put(115,15){$\oplus$}
\put(140,15){\circle*{5}}
\put(140,15){\line(60,0){40}}
\put(178,19){$s_{2}^{\prime }$}
\put(180,15){\circle*{5}}
\end{picture}
\end{center}
\caption{The Coxeter-Dynkin diagram of $SU(3)\times SU(3)$. Here $\bar{t}=%
\frac{1}{2}(1-e_{1}-e_{2}-e_{3})$ , $s\prime _{1}=\frac{1}{2}(e_{1}-e_{2})$
and $s\prime _{2}=\frac{1}{2}(e_{2}-e_{3})$.}
\label{fig:three}
\end{figure}
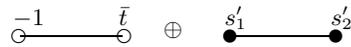

Note that one of the $SU(3)$ is represented by the short roots. The non zero
roots of $SU(3)xSU(3)$ are given by%
\begin{equation}
\pm 1,\pm t,\pm \bar{t},\pm s_{1}^{\prime },\pm s_{2}^{\prime
},s_{3}^{\prime }  \label{e20}
\end{equation}%
where $s_{3}^{\prime }=\frac{1}{2}(e_{3}-e_{1})$ .Using the standard
technique one can form the elements of $W(SU(3))\times W(SU(3))$ of order 36
which is the direct product of two symmetric group $S_{3}$. A further
symmetry is the diagram automorphism of $SU(3)\times SU(3)$ which can be
made by an element $c=[1,\frac{1}{\sqrt{2}}(e_{1}-e_{2})]$ which permutes
the simple roots and preserve the Cartan matrix of the algebra $SU(3)xSU(3)$%
. An extension of the Weyl group $W(SU(3))\times W(SU(3))$ by the element $%
c=[1,\frac{1}{\sqrt{2}}(e_{1}-e_{2})]$ leads to, up to conjugation, the
group $Aut(SU(3)xSU(3))\approx \lbrack W(SU(3))\times W(SU(3))]:Z_{4}$ \cite%
{24} where $Z_{4}$ is the cyclic group of order 4 generated by the element $%
c $. The set of elements can be represented by $[p,q]\oplus \lbrack
p,q]^{\ast }$ where $p,q$ take arbitrary values from the set of scaled roots 
$p,q\in {}\left\{ \pm 1,\pm t,\pm \bar{t},\pm s_{1},\pm s_{2},\pm
s_{3}\right\} $ where $s_{i}=\sqrt{2}s\prime _{i}(i=1,2,3)$.

\section{$SO(8)$ and its subgroups}

The $SO(8)$ algebra plays a special role when embedding in $F_{4}$ since the
long roots of $F_{4}$ are the roots of $SO(8)$. Its Coxeter-Dynkin diagram
illustrates the triality in terms of the cyclic symmetry of the quaternionic
imaginary units:

\begin{figure}[h]
\begin{center}
\begin{picture}(300,120)(0,0)
\put(75,15){$e_{1}$}
\put(90,15){\circle{5}}
\put(91,15){\line(60,0){39}}
\put(140,15){$\frac{1}{2} (1-e_{1} -e_{2} -e_{3} )$}
\put(130,15){\circle{5}}
\put(131,16){\line(2,1){39}}
\put(175,35){$e_{2}$}
\put(170,35){\circle{5}}
\put(129,14){\line(2,-1){39}}
\put(175,-5){$e_{3}$}
\put(170,-5){\circle{5}}
\end{picture}
\end{center}
\caption{The Coxeter-Dynkin diagram of $SO(8)$ .}
\label{fig:four}
\end{figure}
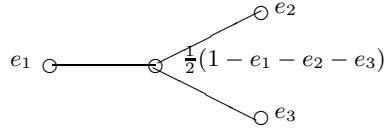

The Weyl group $W(SO(8))$ is represented by the set of elements (\ref{e14a})
and the $Aut(S(O(8))$ is isomorphic to the Weyl group $W(F_{4})$. Since in
reference \cite{12} we have worked $SO(8)$ in some detail here we will deal
with its two special subgroups $SO(7)$ and $G_{2}$.

\subsubsection{$SO(7)$}

The $SO(7)$ diagram can be obtained from that of $SO(8)$ by folding two
branches and averaging the corresponding simple roots \cite{17}.

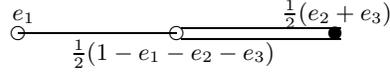
\begin{figure}[h]
\begin{center}
\begin{picture}(300,120)(0,0)
\put(88,20){$e_{1}$}
\put(90,15){\circle{5}}
\put(90,15){\line(60,0){60}}
\put(150,15){\circle{5}}
\put(110,5){$\frac{1}{2} (1-e_{1} -e_{2} -e_{3} )$}
\put(152,17){\line(60,0){60}}
\put(152,13){\line(60,0){60}}
\put(190,20){$\frac{1}{2} (e_{2} +e_{3} )$}
\put(210,15){\circle*{5}}
\end{picture}
\end{center}
\caption{The Coxeter-Dynkin diagram of $SO(7)$ .}
\label{fig:five}
\end{figure}

Figure 5. The Coxeter-Dynkin diagram of $SO(7)$ .

Denote by the reflection generators $r_{1},r_{2},r_{3}$ of $SO(7)$
corresponding to the simple roots $e_{1}$, $\bar{t}=\frac{1}{2}%
(1-e_{1}-e_{2}-e_{3})$ , $\frac{1}{2}(e_{2}+e_{3})$ respectively which can
be expressed as%
\begin{equation}
r_{1}=[e_{1},-e_{1}]^{\ast },r_{2}=[\bar{t},-\bar{t}]^{\ast },r_{3}=[\frac{1%
}{\sqrt{2}}(e_{2}+e_{3}),-\frac{1}{\sqrt{2}}(e_{2}+e_{3})]^{\ast }
\label{e21}
\end{equation}

One can also express $r_{3}$ in terms of the simple roots of $SO(8)$ as the
product of reflection generators corresponding to $e_{2}$ and $e_{3}$ rather
than the one in (\ref{e21}). That would give us $r_{3}=[e_{2},-e_{2}]^{\ast
}[e_{3},-e_{3}]^{\ast }=[e_{1},-e_{1}]$ which gives the same result when
acting on the roots of $SO(7)$ . If we define $d_{1}=\frac{1}{\sqrt{2}}%
(e_{2}-e_{3})$ then we can write the $W(SO(7))$ generators as%
\begin{equation}
r_{1}=[e_{1},-d_{1}\bar{e}_{1}\bar{d}_{1}]^{\ast },r_{2}=[\bar{t},-d_{1}t%
\bar{d}_{1}]^{\ast },r_{1}=[e_{1},-d_{1}\bar{e}_{1}\bar{d}_{1}]  \label{e22}
\end{equation}%
The generators $a=r_{1}r_{2}$ and $b=(r_{1}r_{2}r_{3})^{2}$ satisfy the
generation relation%
\begin{equation}
a^{3}=b^{3}=(ab)^{2}=[1,1]  \label{e23}
\end{equation}%
which is the generation relation of the tedrahedral group of order 12
isomorphic to the group $A_{4}$ of the even permutations of four letters %
\cite{23}. The group elements can be written as%
\begin{equation}
\lbrack p,d_{1}\bar{p}\bar{d}_{1}],p\in T.  \label{e24}
\end{equation}

One can check that the elements%
\begin{equation}
\lbrack 1,-1],[1,\pm 1]^{\ast }  \label{e25}
\end{equation}%
preserve the simple roots by conjugation. This means that the tetrahedral
group in (\ref{e24}) can be extended by the elements in (\ref{e25}) so that
the whole set of elements will read%
\begin{equation}
\lbrack p,\pm d_{1}\bar{p}\bar{d}_{1}],[p,\pm d_{1}\bar{p}\bar{d}_{1}]^{\ast
}.  \label{e26}
\end{equation}%
We note that the set of elements%
\begin{equation}
\lbrack p,d_{1}\bar{p}\bar{d}_{1}],[p,d_{1}\bar{p}\bar{d}_{1}]^{\ast }
\label{e27}
\end{equation}%
form a group isomorphic to the octahedral group $S_{4}$ . The element $%
[1,-1] $ commutes with the elements of $S_{4}$ in (\ref{e27}). Therefore the
set of elements represent a group isomorphic to the group $S_{4}\times Z_{2}$
which is the Weyl group $W(SO(7))$ of order 48. This is the group isomorphic
to $W(SP(3))$ represented by (\ref{e19a}-\ref{e19b}). We could have
different foldings of $SO(8)$ diagram other than the one shown in figure 5.
This would lead to replacing the quaternion $d_{1}$ in (\ref{e26}) by $d_{2}=%
\frac{1}{2}(e_{3}-e_{1})$ and $d_{3}=\frac{1}{2}(e_{1}-e_{2})$ . By
replacing $d_{1}$ by $d_{2}$ and $d_{3}$ in (\ref{e26}) we obtain three
different embeddings of $SO(7)$ in $SO(8)$ . When we stick to the
representation of $W(SO(7))$ in (\ref{e26}) we can show that the 24 non-zero
roots of $SO(8)$ can be decomposed as%
\begin{equation}
\pm 1,\pm e_{1},e_{2},-e_{3},\frac{1}{2}(\pm 1\pm e_{1}\pm (e_{2}+e_{3})),%
\frac{1}{2}(\pm 1\pm e_{1}+(e_{2}-e_{3}))  \label{e28}
\end{equation}%
which represent 18 non-zero roots of $SO(7)$ and the remaining ones are the
6 non-zero weights of the 7-dimensional representation of $SO(7)$

\begin{equation}
-e_{2},e_{3},\frac{1}{2}(\pm 1\pm e_{1}-(e_{2}-e_{3})).  \label{e29}
\end{equation}%
Three different embeddings of $SO(7)$ in $SO(8)$ can be realized by
permuting the indices $(1,2,3)$ in (\ref{e28}-\ref{e29})in the cyclic order.

\subsubsection{$G_{2}$}

The Coxeter --Dynkin diagram of $G_{2}$ can be obtained from that of $SO(8)$
by folding three branches and taking the average of the outer simple roots %
\cite{17}.

\begin{figure}[h]
\begin{center}
\begin{picture}(300,120)(0,0)
\put(30,15){$\frac{1}{2}(1-e_{1}-e_{2}-e_{3})$}
\put(120,15){\circle{7}}
\put(120,18){\line(60,0){60}}
\put(122,15){\line(60,0){59}}
\put(120,12){\line(60,0){60}}
\put(180,15){\circle*{7}}
\put(185,15){$\frac{1}{3} (e_{1} +e_{2} +e_{3} )$}
\end{picture}
\end{center}
\caption{The Coxeter-Dynkin diagram of $G_{2}$.}
\label{fig:six}
\end{figure}

Let us denote by $I=\frac{1}{\sqrt{3}}(e_{1}+e_{2}+e_{3})$ with $I^{2}=-1$.
The simple roots scaled by $\sqrt{2}$ are given by $\alpha _{1}=\frac{1}{2}%
(1-\sqrt{3}I)=e^{-\frac{\pi }{3}I}$ , $\alpha _{2}=\frac{I}{\sqrt{3}}$ and
the reflection generators read $r_{1}=[e^{-\frac{\pi }{3}I},-e^{-\frac{\pi I%
}{3}}]^{\ast }$ , $r_{2}=[I,-I]^{\ast }$. The group $W(G_{2})$ generated by $%
r_{1}$ and $r_{2}$ is the dihedral group $D_{6}$ of order 12. One can obtain
the 12 non-zero roots of $G_{2}$ by acting the generators $r_{1}$ and $r_{2}$
on the simple roots. The weights of the 7-dimensional representation can be
obtained from the highest weight $\sqrt{\frac{2}{3}}e^{\frac{\pi }{6}I}$.

A remark is in order. We can summarize the discussion in this section that
the $SO(7)$ can be embedded in $SO(8)$ triply symmetric way and the $G_{2}$
takes place in the intersection of these three $SO(7)$ in $SO(8)$ \cite{25}

\begin{figure}[h]
\begin{center}
\begin{picture}(300,120)(0,0)
\put(50,40){$SO(7)^{1}$}
\put(60,50){\line(2,1){60}}
\put(60,30){\line(2,-1){60}}
\put(160,80){\line(2,-1){60}}
\put(160,0){\line(2,1){60}}
\put(215,40){$SO(7)^{3}$}
\put(130,80){$SO(8)$}
\put(130,0){$G_{2}$}
\put(140,10){\line(0,60){20}}
\put(140,60){\line(0,60){20}}
\put(130,40){$SO(7)^{2}$}
\end{picture}
\end{center}
\caption{Three-fold embedding of $SO(7)$ in $SO(8)$ .}
\label{fig:seven}
\end{figure}
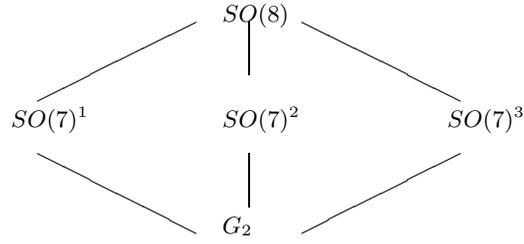

\section{Non-parabolic Subgroups of $Aut(F_{4})$}

So far we have discussed the parabolic subgroups of $Aut(F_{4} )$ related
with the Lie sub-algebras of $F_{4} $ . As we have mentioned before the $%
Aut(F_{4} )$ is the largest crystallographic group in 4-dimensions and
deserves further analysis regarding its chain decomposition through its
maximal subgroups which could be useful for the crystallography in 4-
dimensions. First we discuss the maximal subgroups of $Aut(F_{4} )$ . We
will give the group elements in terms of quaternions and distinguish the
groups by their orders and conjugacy classes. The group orders and conjugacy
classes are not sufficient to understand the group structures. Since we will
write down the group elements explicitly in terms of quaternions the
distinguishing the groups of the same order will not create a problem.
Nevertheless we will denote a group of interest with its order together with
its conjugacy classes in a parenthesis and display the group elements in
terms of quaternions. For example, the group $Aut(F_{4} )$ , being of order
2304 with 29 conjugacy classes will be shortly denoted by 2304(29) and its
quaternionic representation will follow the group notation. We know that
this is not a proper group notation; it should rather have a decomposition
involving invariant subgroups. Since we denote each group by their elements
the order with the conjugacy classes would be sufficient .

\subsection{Maximal Subgroups of $Aut(F_{4})$}

We have three maximal subgroups of $Aut(F_{4} )$ of order 1152.

\subsubsection{A : $W(F_{4})$\ of order 1152(25)}

It is a subgroup of $O(4)$ . We have discussed this group in details which
was represented by the quaternions in (\ref{e11}) :%
\begin{equation*}
W(F_{4})={[T,T]\oplus \lbrack T}^{\prime }{,T}^{\prime }{]\oplus \lbrack
T,T]^{\ast }\oplus \lbrack T}^{\prime }{,T}^{\prime }{]^{\ast }.}
\end{equation*}%
Note that the group $W(F_{4})$ is invariant under the transformation $%
T\leftrightarrow T^{\prime }$ .

\subsubsection{B : The group 1152(19)}

It is a subgroup of $O(4)$ and its quaternionic structure can be written as
follows%
\begin{equation}
{\lbrack T,T]\oplus \lbrack T}^{\prime }{,T}^{\prime }{]\oplus \lbrack T,T}%
^{\prime }{]^{\ast }\oplus \lbrack T}^{\prime }{,T]^{\ast }}  \label{e30}
\end{equation}%
We know that the first two set of elements form a subgroup of order 576. One
can show that the set of elements in (30) is closed by noting that%
\begin{equation}
\lbrack T,T^{\prime }]^{\ast }[T,T^{\prime }]^{\ast }=[T^{\prime },T]^{\ast
}[T^{\prime },T]^{\ast }=[T,T],\quad \lbrack T,T^{\prime }]^{\ast
}[T^{\prime },T]^{\ast }=[T^{\prime },T^{\prime }].  \label{e31}
\end{equation}%
It is clear that it is a maximal subgroup of $Aut(F_{4})$ and will be left
invariant under the transformation $T\leftrightarrow T\prime $ .

\subsubsection{C : The group 1152(34)}

This is the largest crystallographic group in 4-dimensions with proper
rotations . That means it is a finite subgroup of $SO(4)$ .Naturally, it
involves only non-star elements of $Aut(F_{4})$%
\begin{equation}
\lbrack T,T]\oplus \lbrack T^{\prime },T^{\prime }]\oplus \lbrack
T,T^{\prime }]\oplus \lbrack T^{\prime },T].  \label{e32}
\end{equation}

Its closure property is straightforward . Some of its subgroups of order 192
will be of our special interest for they appear as maximal subgroups in some
of the finite subgroups of the Lie group $G_{2}$ \cite{26}. It is also
invariant under the transformation $T\leftrightarrow T^{\prime }$.

Now we discuss, in turn, the maximal subgroups classified under the title A,
B, C.

\subsection{A.The Maximal Subgroups of $W(F_{4})$}

Its parabolic subgroups have been already discussed in section 4. Besides
those groups there are two maximal subgroups of order 576 with the conjugacy
classes 20 and 23. The group 576(23) is the extension of the Weyl group $%
W(SO(8))$ by a cyclic symmetry of the simple roots represented by imaginary
quaternions. The group 576(20) is also a maximal subgroup of the groups
1152(19) and 1152(34).

\subsubsection{A1. The group 576(23).}

It is the extension of the of the group $W(SO(8))$ by a cyclic group of
order 3 and its elements can be written as%
\begin{equation}
\lbrack T,T]\oplus \lbrack T,T]^{\ast }\approx \lbrack T,T]\oplus \lbrack
T^{\prime },T^{\prime }]^{\ast }.  \label{e33}
\end{equation}%
It can be shown that the group can be written as semi-direct product of the
Weyl group of $SO(8)$ and the cyclic group $Z_{3}$,%
\begin{equation*}
W(SO(8)):Z_{3},
\end{equation*}%
where the cyclic symmetry permutes the outer simple roots of $SO(8)$.

\subsubsection{A2. The group 576(20).}

It has the structure%
\begin{equation}
\lbrack T,T]\oplus \lbrack T^{\prime },T^{\prime }].  \label{e34}
\end{equation}

It is also a maximal subgroup in the crystallographic subgroup of $SO(4)$
denoted by 1152(34) and the group 1152(19). No doubt that the elements in (%
\ref{e34}) closes under multiplication. We simply note the non-trivial case,
namely,%
\begin{equation*}
\lbrack T\prime ,T^{\prime }]^{2}=[T,T].
\end{equation*}

\subsection{B.The Maximal subgroups of the group 1152(19)}

\subsubsection{B1. The group 576(20).}

This group is just discussed in A2.

\subsubsection{B2. The group 192(17)}

This group occurs also in the subgroup decomposition of the group 1152(34)
and will be discussed under the subtitle C.

\subsection{C.The Maximal subgroups of the group 1152(34)}

\subsubsection{C1. The group 576(29)}

It has the structure%
\begin{equation}
\lbrack T,T]\oplus \lbrack T,T^{\prime }]\approx \lbrack T,T]\oplus \lbrack
T^{\prime },T]  \label{e35}
\end{equation}

Since the group 576(29) is an index 2 group in the group 1152(34) it should
have two conjugates subgroups which is reflected in the isomorphism above.

\subsubsection{C2. The group 384(31)}

We have the following structure of the group%
\begin{equation}
\lbrack V_{0},T]\oplus \lbrack V_{1},T]\oplus \lbrack V_{0},T^{\prime
}]\oplus \lbrack V_{1},T^{\prime }]  \label{e36}
\end{equation}

This is certainly a maximal subgroup of the group 1152(34) because $%
V_{0}\oplus V_{1}$ form a maximal subgroup of order 16 in the binary
octahedral group $T\oplus T^{\prime }$. It can be embedded in the group
1152(34) triply symmetric way by replacing $V_{1}$ by $V_{2}$ and $V_{3}$ in
(\ref{e36}) in a similar manner where $W(SO(9))$ is embedded in $W(F_{4})$ .

\subsubsection{C3. The group 288(24)}

When we examine the parent group $[O,O]$ we know that the binary octahedral
group $O$ has many maximal subgroups one of which is the dicyclic group(
binary dihedral group) of order 12. It can be generated by two elements $a=%
\frac{1}{2}(1-e_{1}-e_{2}-e_{3})$ and $b=\frac{1}{\sqrt{2}}(e_{1}-e_{2})$
where $a^{6}=b^{4}=1$ satisfying the generation relation $ba^{n}\bar{b}=%
\bar{a}^{n}(n=1,...,6)$. The group can be denoted by $2D_{3}$ where $D_{3}$
is the dihedral group of order 6. When $2D_{3}$ acts on the left and the
binary octahedral group acts on the right we obtain the group 288(24) which
reads in our notation%
\begin{equation}
\lbrack 2D_{3},O]  \label{e37}
\end{equation}%
We can further continue to determine the maximal subgroups of the groups
discussed in the series A,B and C.

\subsubsection{A1. The maximal subgroups of $W(SO(8)):Z_{3}$}

\paragraph{A1.1. The group 288(25)}

This is the group $\lbrack T,T]$ occurring in many groups discussed above.

\paragraph{A1.2. The group $192(13)\approx W(SO(8))$}

It has been discussed before and shown to be the Weyl group of $SO(8)$%
\begin{equation}
\lbrack V_{0},V_{0}]\oplus \lbrack V_{+},V_{+}]\oplus \lbrack
V_{-},V_{-}]\oplus \lbrack V_{0},V_{0}]^{\ast }\oplus \lbrack
V_{+},V_{+}]^{\ast }\oplus \lbrack V_{-},V_{-}]^{\ast }  \label{e38}
\end{equation}%
which is invariant under the cyclic symmetry $Z_{3}$. The action of the
group elements on the hyperoctahedra $V_{0},V_{+},V_{-}$ are as follows:

i) $\lbrack V_{0} ,V_{0} ]$ leaves each hyperoctahedra invariant.

ii) $\lbrack V_{+} ,V_{+} ]$ permutes the three octahedra in the cycylic
order and $\lbrack V_{-} ,V_{-} ]$ does the same in the reverse order.

iii) The element $[V_{i},V_{i}]^{\ast }(i=0,+,-)$ leaves the hyperoctahedron 
$V_{i}$ invariant but interchanges the other two. These properties indicate
that the $[V_{0},V_{0}]$ form an invariant subgroup where the factor group
is the symmetric group of order 6 $\frac{W(SO(8))}{[V_{0},V_{0}]}\approx
S_{3}$.

\paragraph{A1.3. The group $192(16)$}

It can be represented in three equivalent ways and can be proven that they
are the conjugate groups

i) $[V_{0},V_{0}]\oplus \lbrack V_{+},V_{-}]\oplus \lbrack
V_{-},V_{+}]\oplus \lbrack V_{0},V_{0}]^{\ast }\oplus \lbrack
V_{+},V_{-}]^{\ast }\oplus \lbrack V_{-},V_{+}]^{\ast }$

ii) $[V_{0},V_{0}]\oplus \lbrack V_{+},V_{-}]\oplus \lbrack
V_{-},V_{+}]\oplus \lbrack V_{+},V_{+}]^{\ast }\oplus \lbrack
V_{-},V_{0}]^{\ast }\oplus \lbrack V_{0},V_{-}]^{\ast }$

iii) $[V_{0},V_{0}]\oplus \lbrack V_{+},V_{-}]\oplus \lbrack
V_{-},V_{+}]\oplus \lbrack V_{-},V_{-}]^{\ast }\oplus \lbrack
V_{+},V_{0}]^{\ast }\oplus \lbrack V_{0},V_{+}]^{\ast }$

Interestingly enough that each of these conjugate groups leaves one of the
hyperoctahedra invariant. One can easily show that the groups in(i),(ii) and
(iii) leave $V_{0} ,V_{+} $ and $V_{-} $ invariant respectively. Embedding
of the group 192(16) in the group $W(SO(8)):Z_{3} $ follows the cyclic
symmetry of quaternionic units $e_{1} ,e_{2} ,e_{3} $ .

\subsubsection{B1. The Maximal subgroups of the group 576(20)}

\paragraph{B1.1. The group 288(25)}

It has been discussed in A1.1.

\paragraph{B1.2. The group 288(24)}

This group was discussed in C3.

\paragraph{B1.3. The group $192^{\prime }(13)$}

It has the same order and the same number of conjugacy classes with $%
W(SO(8)) $ but not isomorphic to it. It has the structure%
\begin{equation}
\lbrack V_{0},V_{0}]\oplus \lbrack V_{+},V_{+}]\oplus \lbrack
V_{-},V_{-}]\oplus \lbrack V_{1},V_{1}]\oplus \lbrack V_{2},V_{3}]\oplus
\lbrack V_{3},V_{2}].  \label{e40}
\end{equation}

An important difference is that $W(SO(8))$ is a subgroup of $O(4)$ whereas
this group is a subgroup of $SO(4)$. The group $192^{\prime }(13)$ has an
index 6 in the group 1152(34). Its conjugate groups can be obtained by the
conjugation of the element $[V_{+},V_{1}]$ which permutes the 6
hyperoctahedra in the cyclic order $V_{0}\rightarrow V_{3}\rightarrow
V_{-}\rightarrow V_{1}\rightarrow V_{+}\rightarrow V_{2}\rightarrow V_{0}$ .
This would yield the 6 conjugate representations of (\ref{e40}).This group
turns out to be a maximal subgroup of the finite subgroup of $G_{2}$ of
order 1344 preserving the octonion algebra of the set $\pm
e_{i}(i=1,2,...,7) $ \cite{26}.

\subsubsection{C1.Maximal subgroups of the group 576(29)}

All its maximal subgroups which have not been discussed so far also occur as
the maximal subgroups of the group 384(31) and will be discussed below.

\subsubsection{C2.Maximal subgroups of the group 384(31)}

\paragraph{C2.1. The group 192(26)}

It has the structure%
\begin{equation}
\lbrack V_{0},T]\oplus \lbrack V_{1},T]  \label{e41}
\end{equation}

\paragraph{C2.2. The group 192(23)}

It can be represented by%
\begin{equation}
\lbrack V_{0},T]\oplus \lbrack V_{0},T^{\prime }]  \label{e42}
\end{equation}

\paragraph{C2.3. The group 192(20)}

This group has an interesting structure which can be written as%
\begin{equation}
\lbrack a,T]\oplus \lbrack b,T^{\prime }]  \label{e43}
\end{equation}%
where the set of elements of $a$ is generated by $\frac{1}{\sqrt{2}}\left(
1+e_{1}\right) $ and the set $b=e_{3}a$. The set $[a,T]$ forms an invariant
subgroup of order 96. The set of elements $a$ and $b$ generate a dicyclic
group of order 16 as we discussed before however as $a$ and $b$ are paired
with different subsets of the binary octahedral group the dicyclic group is
not a subgroup of the group 192(20). The set of elements of a and b are
given by%
\begin{equation}
a=\left\{ \pm 1,\pm e_{1},\frac{1}{\sqrt{2}}\left( \pm 1\pm e_{1}\right)
\right\} ,\quad b=\left\{ \pm e_{2},\pm e_{3},\frac{1}{\sqrt{2}}\left( \pm
e_{2}\pm e_{3}\right) \right\}  \label{e44}
\end{equation}

\paragraph{C2.4. The group 192(17)}

It can be represented by%
\begin{equation}
\lbrack V_{0},T]\oplus \lbrack V_{1},T^{\prime }]  \label{e45}
\end{equation}%
which is also a subgroup of the group 576(20). It is one of the maximal
subgroup of the finite subgroup of the Lie group of $G_{2}$ order 12096
which leaves the quaternion decomposition of the octonionic root system of
the exceptional Lie algebra $E_{7}$ \cite{26} invariant.

\section{Conclusion}

The automorphism group $Aut(F_{4})$ of the root system of the exceptional
Lie algebra $F_{4}$ is the largest crystallographic group in 4-dimensions
which has not been discussed in the literature using quaternions. This work
not only relates this crystallographic group to the Coxeter- Dynkin diagram
of $F_{4}$\ but also discusses its relevance to other Lie algebraic
structures as well as to the 4-dimensional Euclidean geometry. We have
discussed the decomposition of $Aut(F_{4})$ down to the groups of order 192
and shown that a number of groups of order 192 have different structures
related to different geometries. It is perhaps also interesting to continue
the same decomposition to determine the groups acting in 3-dimensions . In
this context we have discussed only the Weyl groups $W(S(7))\approx W(SP(3))$%
.

We have noted that two groups of order 192, namely, the groups $192^{\prime
}(13)$ and 192(17) occur as maximal subgroups in the finite subgroups of the
Lie group $G_{2}$. The group $192^{\prime }(13)$\ is a maximal subgroup of a
group $Z_{2}^{3}\cdot PSL_{2}(7)$ of order 1344 which is a finite subgroup
of $G_{2}$ preserving the set of imaginary octonions $\pm e_{i}(i,1,2,\cdots
,7)$ \cite{16}. Here $PSL_{2}(7)$\ is the famous Klein's simple group of
order 168 and $Z_{2}^{3}=Z_{2}\times Z_{2}\times Z_{2}$ is the elementary
abelian group of order 8. The group 192(17) is the maximal subgroup of the
Chevalley group $G_{2}(2)$ of order 12096 which leaves the octonionic roots
of the Lie algebra $E_{7}$ invariant .

The Weyl group $W(SO(8))$ which is the group 192(13) is also a maximal
subgroup of a group $Z_{2}^{3}:PSL_{2}(7)$ of order 1344 which is, in turn,
a maximal subgroup of the simple group $A_{8}$, even permutations of 8
letters. The group $A_{8}$\ is related to the Weyl group $W(E_{7})$ through $%
W(SU(8)$ and is a maximal subgroup of the Chevalley group $SO_{7}(2)$ \cite%
{27}. It is also interesting to note that some finite subgroups of $SO(4)$
also occur in the phase transitions of the liquid helium $^{3}He$ \cite{28}.

We believe that the group structures and their quaternionic representations
will be useful in various fields of physics which may need the finite
subgroups of $O(4)$.


\begin{thebibliography}{99}
\bibitem{not1} We use the notation of reference \cite{11}
\bibitem{not2} The notation is such that $A:B$ is the semi-direct product of
two groups $A$ and $B$ where $A$ is the invariant subgroup of the product
group \cite{22}.
\end{thebibliography}
\end{document}